\documentclass[5p,numbers,sort&compress]{elsarticle}
\usepackage{epsfig,bm,epsf,graphics}
\usepackage{amsmath}
\begin{document}
\begin{frontmatter}
\title{Monte Carlo Study of the Spin Transport in Magnetic Materials}

\author[UCP]{Y. {\sc Magnin}}
\author[UCP,OU]{K. {\sc Akabli}}
\author[UCP]{H. T. {\sc Diep}\corref{corresponding_author}}
\ead{diep@u-cergy.fr}
\cortext[corresponding_author]{To whom correspondence should be addressed.}
\author[OU]{I. {\sc Harada}}

\address[UCP]{Laboratoire de Physique Th\'eorique et Mod\'elisation,
\\
CNRS UMR 8089, Universit\'e de Cergy-Pontoise,
\\
2, avenue Adolphe Chauvin, F-95302 Cergy-Pontoise Cedex, France.}

\address[OU]{ Graduate School of Natural Science and Technology, Okayama University\\
3-1-1 Tsushima-naka, Kita-ku, Okayama 700-8530, Japan..}

  \begin{abstract}
The resistivity in magnetic materials has been theoretically shown to depend on the spin-spin correlation function which in turn depends on the magnetic-field, the density of conduction electron, the magnetic ordering stability, etc.  However, these theories involved a lot of approximations, so their validity remained to be
confirmed.  The purpose of this work is  to show by newly improved extensive Monte Carlo (MC) simulation the resistivity of the spin resistivity from low-$T$ ordered phase to high-$T$ paramagnetic phase in ferromagnetic and antiferromagnetic films.  We take into account the interaction between the itinerant spins and the localized lattice spins as well as the interaction between itinerant spins themselves.    We show that in ferromagnets the resistivity shows a sharp peak at the magnetic phase transition in agreement with previous theories in spite of their numerous approximations.
 Resistivity in antiferromagnets on the other hand shows no  peak for the SC, BCC and diamond lattices.  Discussion on the origin of these resistivity behaviors is given.

\begin{keyword}
spin transport \sep Monte Carlo simulation \sep magnetic resistivity \sep magnetic materials
\PACS 72.25.-b \sep 75.47.-m
    \end{keyword}
  \end{abstract}
\end{frontmatter}

\section{Introduction}
The magnetic
resistivity has been extensively studied by both theories and experiments
in the last fifty years. Experiments have shown that the resistivity
indeed depends on the itinerant spin
orientation and the lattice spin ordering \cite{Fert-Campbell,Shwerer,Stishov,Stishov2,Matsukura,Brucas}.
At low temperature ($T$), the main magnetic scattering is due to spin-wave excitations \cite{Kasuya,Turov}.  The resistivity is proportional to $T^2$.  However at higher $T$ the spin-wave theory is not valid,
such a calculation of the  resistivity is not possible, in particular in the critical region around the Curie temperature $T_c$ in simple ferromagnets, let alone other complicated magnetic orderings.  Experiments on various magnetic
materials have found in particular an anomalous behavior of the
resistivity at the critical temperature where  the system
undergoes the ferromagnetic-paramagnetic phase
transition \cite{Shwerer,Stishov,Stishov2,Matsukura,Brucas}.
Very recent experiments such as those performed on ferromagnetic SrRuO$_3$ thin films\cite{Xia}, superconducting BaFe$_2$As$_2$ single crystals\cite{Wang-Chen}, La$_{1-x}$Sr$_x$MnO$_3$\cite{Santos}, Mn$_{1-x}$Cr$_x$Te\cite{Li} and other compounds\cite{Lu,Du,Zhang,McGuire} show different forms of anomaly of the magnetic resistivity at the transition temperature.
de Gennes and
Friedel's first explanation in 1958\cite{DeGennes} for the resistivity behavior near $T_c$ was based
on the interaction between the spins of conduction electrons and the
lattice spins.
The resistivity was thus expected to depend  strongly on the
spin ordering of the system.
They have suggested that the magnetic
resistivity is proportional to the spin-spin correlation, therefore it should behave as the magnetic
susceptibility with a divergence at $T_c$ due to "long-range" fluctuations of the magnetization. Other authors \cite{Craig,Fisher,Alexander} subsequently suggested that the shape of the resistivity results mainly
from "short-range" correlation at $T\geq T_c$. Fisher and
Langer \cite{Fisher} have shown in particular that the form of the resistivity
cusp depends on the correlation range.  To see more details on the role of the spin-spin
correlation, we quote a work by Haas\cite{Haas} and a more recent work
of Kataoka\cite{Kataoka} where  the spin-spin
correlation function has been calculated.  Recently, Zarand et al \cite{Zarand}
have used the picture that the itinerant spin is mainly scattered by impurities which are characterized by a "localization length" in the sense of Anderson's localization.  They found that the peak's height depends on this localization length.  Note that since the
giant magnetoresistance (GMR) was discovered experimentally twenty
years ago in magnetic multilayers \cite{Baibich,Grunberg},  intensive
investigations on the spin resistivity, both experimentally and theoretically, have been
carried out\cite{Fert,review}. The "spintronics" was
born with a spectacular rapid development in relation with
industrial applications.  For  recent overviews, the reader is
referred to Refs. \cite{Dietl} and \cite{Barthe}.  In spite of these intensive
investigations, except our works\cite{Akabli,Akabli2}, there have been no Monte Carlo
(MC) simulations  performed regarding the temperature
dependence of the spin transport.
In these works , we have investigated by MC
simulations the effects  of magnetic ordering on the spin current
in magnetic multilayers.

 In this paper we improve our previous MC simulations to study the
transport of itinerant electrons in ferromagnetic and antiferromagnetic crystals.
We use the Ising model and take into account
interactions between lattice spins and itinerant spins. We show
that in ferromagnets we obtain, with our new MC averaging method, much better results for the magnetic resistivity which shows a huge peak at the transition temperature
while in antiferromagnets, the resistivity does not
show such a peak.

 The paper is organized as follows. Section 2 is devoted
to the description of our model and the rules that govern its
dynamics.
In section 3, we describe our MC method and discuss the results
we obtained for ferromagnets. Results on antiferromagnets are shown in Section 4.
Concluding remarks are given in Section 5.

\section{Model}

\subsection{Interactions}
We consider a thin film of FCC structure with two symmetrical (001) surfaces. The total number of cells
is $N_x\times N_y \times N_z$ where each cell has four spins.  Spins localized at
FCC lattice sites are called "lattice spins"
 hereafter.  They interact with each other
through the following Hamiltonian

\begin{equation}
\mathcal H_l=-J\sum_{\left<i,j\right>}\mathbf S_i\cdot\mathbf S_j,
\label{eqn:hamil1}
\end{equation}
where $\mathbf S_i$ is the Ising spin at lattice site $i$,
$\sum_{\left<i,j\right>}$ indicates the sum over every
nearest-neighbor (NN) spin pair $(\mathbf S_i, \mathbf S_j)$, $J$ being the NN interaction.
We consider in this paper both $J>0$ (ferromagnets) and $J<0$ (antiferromagnets).

We consider a flow of itinerant spins interacting with each other and
with the lattice spins.  The interaction between itinerant spins is
defined by

\begin{equation}
\mathcal H_m=-\sum_{\left<i,j\right>}K_{i,j}\mathbf
s_i\cdot\mathbf s_j,  \label{eqn:hamil2}
\end{equation}
where $\mathbf s_i$ is the itinerant Ising spin at position  $\vec
r_i$, and $\sum_{\left<i,j\right>}$ denotes a sum over every spin
pair $(\mathbf s_i, \mathbf s_j)$.  The interaction $K_{i,j}$
depends on the distance between the two spins, i.e. $r_{ij}=|\mathbf
r_i-\mathbf r_j|$.  A specific form of $K_{i,j}$ will be chosen
below.  The interaction between itinerant spins and lattice spins
is given by

\begin{equation}
\mathcal H_r=-\sum_{\left<i,j\right>}I_{i,j}\mathbf
s_i\cdot\mathbf S_j,  \label{eqn:hamil3}
\end{equation}
where the interaction $I_{i,j}$ depends on the distance
between the itinerant spin $\mathbf s_i$ and the lattice spin
$\mathbf S_i$. For the sake of simplicity, we assume the same form
for $K_{i,j}$ and $I_{i,j}$, namely,
\begin{eqnarray}
K_{i,j}&=& K_0\exp(-r_{ij})\label{eqn:hamil5}\\
I_{i,j}&=& I_0\exp(-r_{ij})\label{eqn:hamil6}
\end{eqnarray}
where $K_0$ and $I_0$ are constants.

\subsection{Monte Carlo Method}

Before calculating the resistivity, we determine the critical
temperature $T_c$ below which the system is in the ordered phase using Eq. (\ref {eqn:hamil1}). To this end,
we perform standard Metropolis MC simulations to determine various physical quantities at different $T$\cite{Binder}.

Once the lattice is equilibrated at a given $T$, we inject $N_0$
itinerant spins into the system.  The itinerant spins move in the $x$ direction under the effect of
an electric field.
We use the periodic boundary conditions to ensure that the average density of
itinerant spins remains constant with evolving time (stationary
regime).

Note that unlike in the previous works\cite{Akabli,Akabli2} where the lattice spin
configuration is frozen while calculating the resistivity,
we use here several thousands of configurations in each of which the resistivity is
averaged with many thousands of passages.  In our previous works, though the overall number of MC steps per spin was as high as in this work, the fact that we have used only a dozen lattice configurations for resistivity calculation has shown strong fluctuations in the result.  In this work, we have made a new device in two steps:

i) For each lattice configuration all itinerant spins move through the system during typically one thousand MC steps.  Then we thermalize again the lattice for several thousands of MC steps before continuing the averaging of the resistivity for another thousand MC steps per spin.  We repeat this cycle for 200 times. In doing so, each itinerant spin was averaged over $10^6$ MC step using "uncorrelated" 200 lattice configurations in all.

ii) The resistivity $R$ is defined as
$R=\frac{1}{n}$
where $n$ is the number of itinerant spins crossing a unit area
perpendicular to the $x$ direction per unit of MC time. To know this number, we count them at three  "detector" surfaces perpendicular to the $x$ direction: the first at $N_x/4$, the second at $N_x/2$ and the third at $3N_x/4$.  Averaging the resistivity at these three system positions helps to improve further the results (in our previous works\cite{Akabli2} we counted them only at the end of the sample).

As will be shown below, these extensive
configuration and space averages give much better results with respect to those in previous works.
The dynamics of itinerant spins is governed by the
following interactions:

i) an electric field $\mathbf E$ which is applied in the $x$ direction.
Its energy is given by
\begin{equation}
\mathcal {H}_E=-e\mathbf E \cdot \mathbf r_i,
\end{equation}
where $ \mathbf r_i$ is the distance traveled by the itinerant spin $\mathbf s_i$ in a MC step, $e$ its
charge.  The orientation of $ \mathbf r_i$ is taken at random, its magnitude is taken from a uniform  distribution between 0 and $r_0$ where $r_0$ is the nearest-neighbor distance;

ii) a chemical potential term ("concentration gradient" effect) given by
\begin{equation}
\mathcal {H}_{c}= Dn(\mathbf r),
\end{equation}
where $n(\mathbf r)$ is the concentration of
itinerant spins in a sphere of radius $D_2$ centered at $\mathbf r$.
$D$ is a constant taken equal to $K_0$ for simplicity;

iii) interactions between a given itinerant spin and lattice spins
inside a sphere of radius $D_1$ (Eq.~\ref{eqn:hamil3});

iv) interactions between a given itinerant spin and other
itinerant spins inside a sphere of radius $D_2$
(Eq.~\ref{eqn:hamil2}).

The
simulation is carried out as follows: at a given $T$ we calculate
the energy of an itinerant spin $\mathbf s_i$ by taking into account all the
interactions described above.  Then we tentatively move the spin
under consideration to a new position with a step $\mathbf r_i$ in
an arbitrary direction. Note that this move is immediately rejected
if the new position is inside a sphere of radius $\Delta_0$ centered at a
lattice spin or an itinerant spin. This excluded space represents the
Pauli exclusion principle in the one hand, and the interaction with
lattice phonons on the other hand.  If the new position does not lie
in a forbidden region of space, then the move is accepted with a
probability given by the standard Metropolis algorithm\cite{Binder}.

\section{Results on Ferromagnetic Thin Films}

We let $N_0$ itinerant spins travel through the system several
thousands of passages until a steady state is reached before averaging the spin resistivity.

The parameters we
use in our calculations for ferromagnets are $s=S=1$ and $N_x= N_y=20$ and
$N_z= 8$. Other parameters are $D_1=D_2=1$ (all distances are in unit of the FCC
cell length), $K_0=I_0=0.5$ and $D=0.5$ unless otherwise stated, $N_0=8\times 20^2$ (namely
one itinerant spin per FCC unit cell), $\Delta_0=0.05$ and $r_0=\sqrt{2}/2$, the FCC nearest-neighbor distance.
At each
$T$ the equilibration time for the lattice spins lies around
$10^6$ MC steps per spin and we compute statistical averages for the resistivity over
$10^6$ MC steps per spin.
Note that, as described in subsection 2.2, in this work the averaging length $10^6$ MC steps per spin has been divided
into 200 segments of 5000 MC steps; between two consecutive segments we thermalize again our lattice over several thousands of MC steps par spin to explore a maximum of lattice configurations encountered by itinerant spins.  In doing so we reduce statistical fluctuations observed in our old works\cite{Akabli,Akabli2}.

\begin{figure}[htp!]
\centering
\includegraphics[width=40mm,height=80mm,angle=-90]{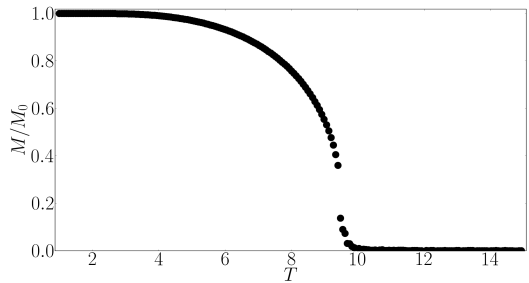}
\caption{Lattice magnetization versus temperature $T$ for $N_z=8$, $N_x=N_y=20$.
$T_c$ is $\simeq 9.58$ in unit of $J=1$.}\label{fig:M(T)}
\end{figure}

We show in Fig. \ref{fig:M(T)} the lattice magnetization versus
$T$ for $N_z=8$, $N_x=N_y=20$, $J=1$. We find  $T_c\simeq
9.58$ for the critical temperature of the lattice spins.

\begin{figure}
\includegraphics[width=40mm,height=80mm,angle=-90]{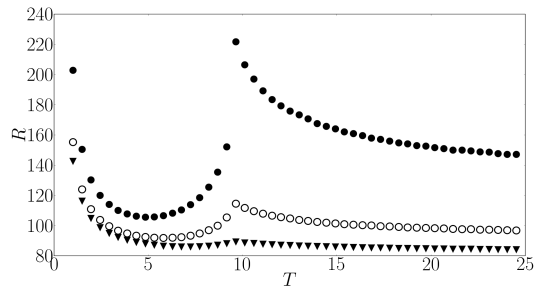}
\caption{Resistivity $R$ in arbitrary unit versus temperature
$T$ for several values of $I_0$: 2 (black circles), 1 (void circles), 0.5 (black triangles).  Other parameters: $N_x=N_y=20$, $N_z$=8, $E=1$, $K_0=0.5$, $D=0.5$, $D_1=1$. }\label{Rbis}
\end{figure}

We show in Fig. \ref{Rbis} the resistivity versus $T$ without magnetic field for several values of $I_0$,  interaction between itinerant spins and lattice spins.
Several remarks are in order:

i) As seen here, no significant fluctuations of the data are observed in the whole temperature range, thanks to
our new averaging device (see subsection 2.2).

ii) At $T_c$, $R$ exhibits a peak at the transition temperature.  The height of the peak decreases with decreasing $I_0$. We see thus that the peak is a consequence of the interaction between itinerant spins and lattice spins.  The resistivity shows almost no peak for $I_0=0.5$.  This case corresponds to a metal where the interaction between itinerant and lattice spins is very weak.

iii) We can
explain the existence of the peak by the following argument:
 the peak is due to the coupling through $I_0$ of itinerant spins to the fluctuations of the lattice spins in the critical region around
$T_c$.  In our recent work, we found
 from our MC simulation\cite{Akabli2}  that the resistivity's peak is due to
 the scattering by antiparallel-spin clusters
 which exist when
 one enters the critical region.
 Below the transition
temperature, there exists a single large cluster of lattice spins
with some isolated "defects" (i. e. clusters of antiparallel
spins), so that the resistance decreases with decreasing $T$ just after $T_c$.

iv) However, at very low $T$,
the resistivity increases with decreasing $T$.   The origin of this behavior comes from the freezing of
the itinerant spins due to their interaction with lattice spins and with themselves.
This is very similar to the crystallization of interacting particles at low $T$.
We have tested this interpretation by reducing the strength of the interactions $K_0$ and $I_0$. As a matter of fact,
$R$ increases more slowly with decreasing $T$.  This is seen in Fig. \ref{Rbis} at low $T$ where $R$ is smaller for smaller $I_0$. Note that the increase of $R$ at very low $T$ was observed in many experiments on various materials not limited to ferromagnets\cite{Wang-Chen,Santos,Li,Du}.

v) In the paramagnetic phase, as $T$
increases, small clusters will be broken more and more into single disordered
spins, so that there is no more energy barrier between successive
positions of itinerant spins on their trajectory. The resistance,
though high, is thus decreasing with increasing $T$ and saturated as
$T\rightarrow \infty$.

\begin{figure}
\begin{center}
\includegraphics[width=45mm,height=55mm,angle=0]{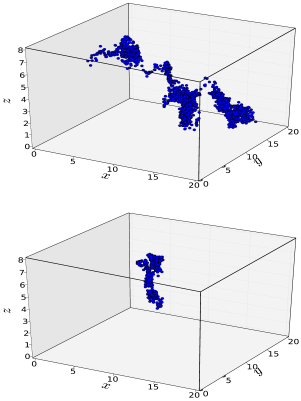}
\caption{ Travel path of an itinerant spin at $T=5$ (upper) and at $T=9.79$ (lower) for $I_0=2$. Other parameters are the same as in Fig. \ref{Rbis}. See text for comments.}\label{travel}
\end{center}
\end{figure}

During the simulation, we have followed one itinerant spin among 3200 and recorded its successive positions. We show in Fig. \ref{travel} its travel path at $T=5$ where $R$ is very low and at $T=9.79$ where $R$ is highest.  As seen, even at $T=5$, the spin spends a lot of time to overcome the scattering by lattice spins.  Almost four thousands of trial moves are needed to get through the system.  Meanwhile, at $T=9.79$, the spin under consideration spends all this time in a small area. Note that at this peak's temperature, very few spins can go until the end unless we increase the electric field.

\subsection{Effect of magnetic field.}

\begin{figure}
\includegraphics[width=40mm,height=80mm,angle=-90]{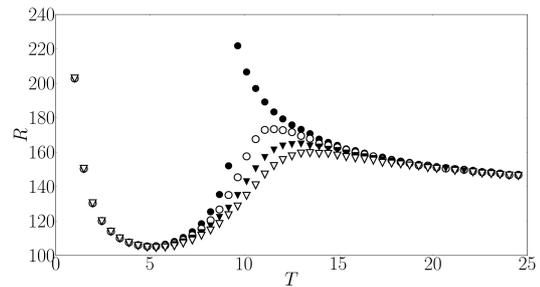}
\caption{Resistivity $R$ in arbitrary unit versus temperature
$T$, for different magnetic fields $B$: 0 (black circles), 0.25 (void circles), 0.5 (black triangles), 0.75 (void triangles). $I_0=2$ and other parameters taken the
same as in Fig. 2.}\label{R_B}
\end{figure}

Kataoka\cite{Kataoka} has shown by a Boltzmann's equation formalism that the magnetic field reduces
the peak's height.
This is what we observed in
simulations.  We show results of $R$ for several fields in Fig.
\ref{R_B}. The peak reduction is stronger for stronger fields. This is easily understood: when a magnetic field is applied on a ferromagnet, the
phase transition is suppressed because the magnetization is not zero at any $T$.
The field reduces critical fluctuations, and hence the number
of clusters of antiparallel spins.  The peak of the resistivity is therefore reduced and disappears at high fields.

To close this section, let us emphasize that in ferromagnets, we have improved the previous results\cite{Akabli2} by using a new averaging procedure.   We found that
the height of the peak is intimately related to the strength of the interaction between itinerant spins and lattice spins (Fig. \ref{Rbis}).

\section{Results on Antiferromagnetic Thin Films}

In the case of antiferromagnets, we study first a film with simple cubic (SC) lattice
structure. This is because the FCC lattice used in the ferromagnetic case shown above
becomes fully frustrated if we use an antiferromagnetic interaction. The frustrated case is very
particular \cite{Magnin}, it cannot be treated on the same footing as the non frustrated case.



Before showing the results on a SC antiferromagnet, let us emphasize the following point.
The picture of defect clusters of down spins embedded in a up-spin sea that we used above to explain the
behavior of the resistivity in ferromagnets should be modified in the case of antiferromagnets: in antiferromagnets defects are domain walls, clusters on the two sides of a wall both have antiferromagnetic
ordering with opposite parity.  An itinerant spin crossing a domain wall does not have the same scattering
as in a ferromagnet.  Its scattering depends on the numbers of up spins and down spins in the sphere of radius $D_1$.  In other words, the scattering depends on the energy landscape in the crystal: the itinerant spin will stay a longer time where its energy is low, and a shorter time where its energy is high.

\begin{figure}
\includegraphics[width=40mm,height=80mm,angle=-90]{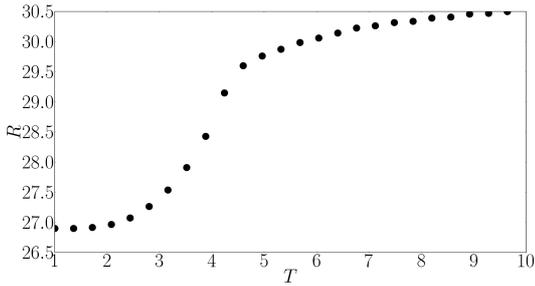}
\caption{ SC AF case. Resistivity $R$ in arbitrary unit versus temperature
$T$, in zero magnetic field, with electric field $E=1$, $I_0=K_0=0.5$.}\label{R-AF}
\end{figure}

\begin{figure}
\includegraphics[width=40mm,height=80mm,angle=-90]{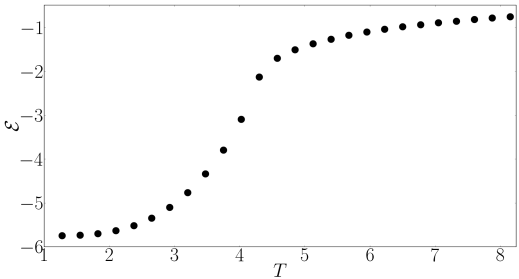}
\caption{ SC AF case. Internal energy $\cal E$ in unit $|J|=1$ versus temperature
$T$, in zero magnetic field, with electric field $E=1$, $I_0=K_0=0.5$.}\label{E-AF}
\end{figure}

The resistivity versus $T$ in zero magnetic field is shown in Fig. \ref{R-AF} with $D_1=D_2=1$.  Several remarks are in order: i) one observes the absence of a peak of $R$; ii) the variation of $R$ with $T$ has the same shape as the internal energy versus $T$  shown in Fig. \ref{E-AF}, therefore $dR/dT$ shows a peak similar to the specific heat.   The peak of $dR/dT$ has been experimentally observed in many materials, in particular in MnSi \cite{Stishov,Stishov2} among others \cite{Shwerer}.

The absence of a peak at $T_N$ observed here certainly comes from the fact that the motion of an itinerant electron is not sharply slowed down at $T_N$ by numerous clusters of opposite spins.   Let us say it again in another manner: the absence of a peak at the transition is due to the
fact that the motion of an itinerant spin depends on its immediate environment: in ferromagnets, the variation of its energy $\Delta E$ going from a "parallel" cluster to a nearby "defect" (or antiparallel) cluster is much larger than the energy variation going from a cluster of antiferromagnetic ordering to a cluster which is a defect but a defect with an antiferromagnetic structure in the SC antiferromagnetic case.  The smaller $\Delta E$ gives rise to a larger spin mobility i. e. a smaller $R$.  Note that experimental data show just a shoulder in antiferromagnetic LaFeAsO\cite{McGuire}.

\begin{figure}
\includegraphics[width=40mm,height=80mm,angle=-90]{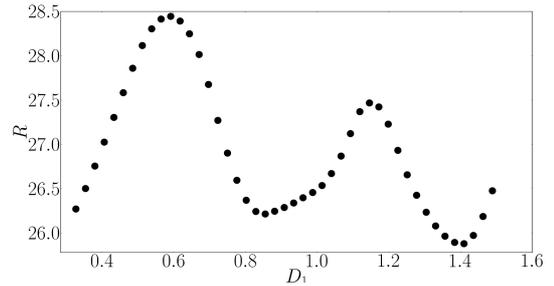}
\caption{ SC AF case. Resistivity $R$ in arbitrary unit versus $D_1$ at $T=1$, in zero magnetic field, with electric field $E=1$, $I_0=K_0=0.5$. }\label{R-D1}
\end{figure}

We show now in Fig. \ref{R-D1} the effect of $D_1$ on the resistivity at a given temperature.  We observe here an oscillatory behavior
of $R$.  By analyzing the ratio of numbers of up spins and down spins in the sphere of radius $D_1$,  we found that this ratio oscillates
with varying $D_1$: the maxima (minima) of $R$ correspond to the largest (smallest) numbers of parallel (antiparallel) spins in the sphere. This finding is consistent with what we said before, namely $R$ is large when the energy of itinerant spin is low (i. e. large number of parallel spins).

\begin{figure}
\includegraphics[width=40mm,height=80mm,angle=-90]{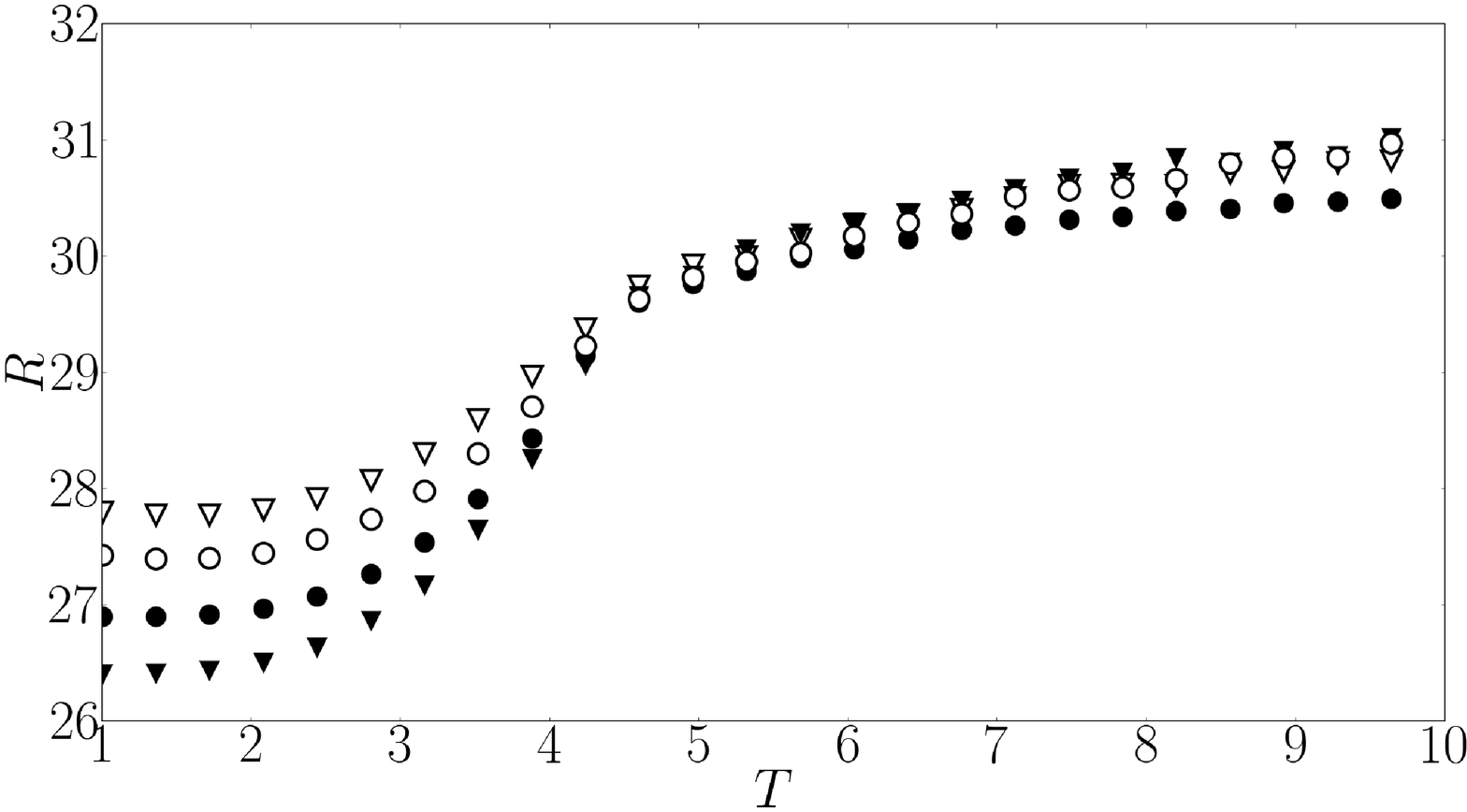}
\caption{ SC AF case. Resistivity $R$ versus $T$ for several values of $D_1$: 1 (black circles), 1.2 (void triangles), 1.4 (black triangles),
  1.6 (void circles), $E=1$, $B=0$, $I_0=K_0=0.5$, $D=0.35$.}\label{R-T-AF}
\end{figure}

We show in Fig. \ref{R-T-AF}  the resistivity versus $T$ for several $D_1$.  As seen here, the change of $R$ at a given $T>T_N$ with varying $D_1$ is much smaller than that for $T<T_N$.  It is interesting to note that
$R$ does not depend on $D_1$ at  $T_N$. Further analysis should be carried out to understand this behavior at the transition point.

\begin{figure}
\includegraphics[width=40mm,height=80mm,angle=-90]{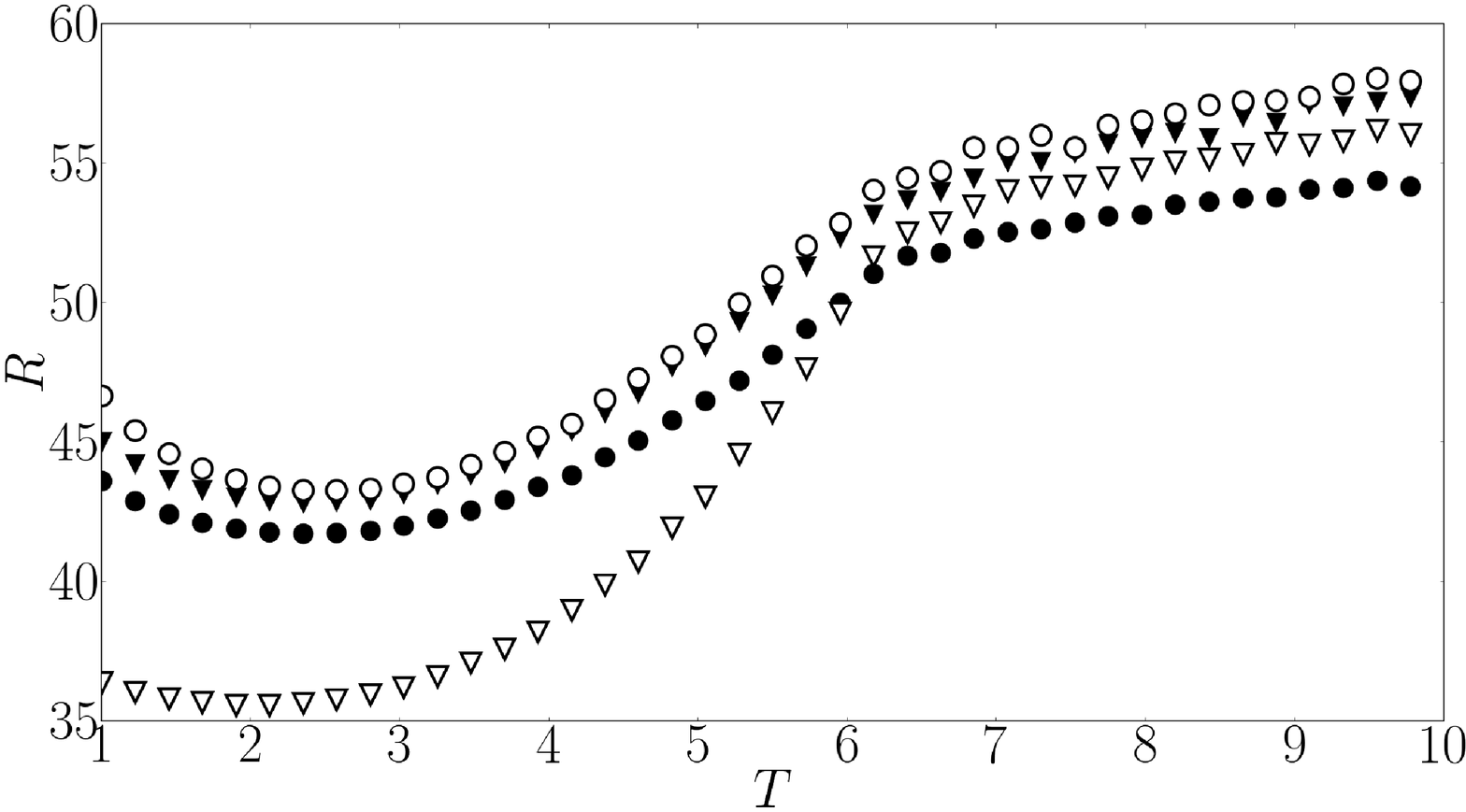}
\caption{ BCC AF case. Resistivity $R$ versus $T$ for several values of $D_1$: 1 (black circles), 1.2 (void triangles), 1.4 (black triangles),
  1.6 (void circles), $E=1$, $B=0$, $I_0=K_0=1$, $D=0.5$.}\label{BCC-AF}
\end{figure}

\begin{figure}
\includegraphics[width=40mm,height=80mm,angle=-90]{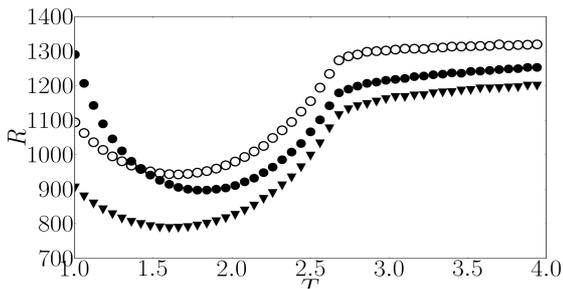}
\caption{ Diamond-lattice AF case. Resistivity $R$ versus $T$ for $D_1$: 0.5 (black triangles), 0.75 (black circles), 1 (void circles).  $E=1$, $B=0$, $I_0=K_0=0.5$, $D=0.35$.}\label{DIAMOND-AF}
\end{figure}

Finally, to compare with the SC AF case, let us show the results of the BCC  and diamond-lattice antiferromagnets in
Figs. \ref{BCC-AF} and \ref{DIAMOND-AF}.  As seen, these cases shows no peak for any value of $D_1$.  Note that $R$ in both cases increases as $T$ decreases to zero, unlike the SC AF case.   The origin of this increase lies in the freezing of itinerant spins at low $T$.  The degree of freezing depends on the lattice structure and on the strength of the interactions of the itinerant spins with their environment as discussed in the ferromagnetic section.

\subsection{Discussion}
In the case of ferromagnets, the coupling of the motion of itinerant spins to the correlation
of the lattice spins gives rise to the peak of the resistivity in the transition region. Depending on the strength of this coupling, the peak can be very sharp or rounded at $T_c$.  The picture of scattering by clusters suggested by us is naturally consistent with the correlation interpretation.

In the case of antiferromagnets, the polarized itinerant spins are coupled to both parallel and antiparallel lattice spins.  Due to the opposite correlation signs, their respective effects are partially canceled out giving rise to an effective coupling  weaker than that in ferromagnets.  It is therefore not surprising that the peak is absent unlike in the ferromagnetic case.

\section{Conclusion}
In this work, we have improved our MC simulations by averaging the resistivity over
a large number of lattice spin configurations.  The results shown above
are much better than those in our previous works \cite{Akabli,Akabli2}. Though the physics
in the ferromagnetic case are not qualitatively altered but the precision on the peak position of $R$
is  excellent and the statistical fluctuations in the paramagnetic phase are reduced.
The spin resistivity is
strongly dependent on the temperature. In ferromagnets, at very low $T$ the itinerant spins are somewhat frozen. As $T$ increases, their motion is thermally activated making a decrease of $R$. However, as $T$ increases further, the system enters the transition region,  $R$ increases and undergoes a huge peak at the
ferromagnetic transition temperature.
At higher temperatures, the lattice spins are disordered, the
resistivity is  still large but it decreases with increasing $T$.    The existence of the peak in ferromagnets is in agreement with theories, in particular those of Zarand et al \cite{Zarand} and Haas \cite{Haas} where interaction between itinerant spins and lattice spins is dominant.

We have also shown here results of some antiferromagnets.  The absence of a peak of $R$ in the SC, BCC and diamond lattices confirms the prediction of Haas \cite{Haas}.
Let us emphasize that our results on frustrated FCC AF \cite{Magnin} and on Heisenberg BCC AF \cite{Akabli3} show that the shape of the resistivity in antiferromagnets strongly depends  on the spin model and the nature, i. e. first or second order, of the lattice phase transition.
The extension of the Boltzmann's theory to the case of antiferromagnets is under way.

KA acknowledges a financial support from the JSPS for his stay at Okayama University.
IH is grateful to the University of Cergy-Pontoise for an invitation during the course of the present work.

{}

\end{document}